# CDP: a Content Discovery Protocol for Mobile P2P Systems

**Taoufik Yeferny, Sofian Hamad and Salem Belhaj**

Northern Border University, Arar - Saudi Arabia

**Summary**
The emergence of affordable wireless and mobile devices was a key step towards deploying mobile peer-to-peer (P2P) systems. The latter allow users to share and search diverse multimedia resources over Mobile Ad-hoc Networks (MANETs). Due to the nature of MANETs, P2P mobile systems brought up many new thriving challenges, in particular with regard to the content discovery issue. Thus, the design of an efficient content discovery protocol has been of paramount importance. The thriving challenge is to (i) locate relevant peers sharing pertinent resources for users' queries and then (ii) to ensure that those peers would be reached by considering different MANET constraints (e.g., peer mobility, battery energy, peer load, to cite a few). Even though the literature witnesses a wealthy number of content discovery protocols, only few of them have considered the above-mentioned requirements. To overcome this shortage, we introduce in this paper an efficient Content Discovery Protocol (CDP) for P2P mobile systems. The idea underlying our proposal is to route users' queries to relevant peers that are (1) more suitable to answer the user query according to its content, (2) more stable to relay query hits, (3) less loaded to avoid the network congestion issue and (4) having more battery lifetime to avoid the network partitioning issue. The performed experiments show that CDP outperforms its competitor in terms of effectiveness and efficiency.

## 1. Introduction

In the last few years, Mobile Ad-hoc Networks (MANETs) have achieved great progress due to their success in different fields. MANETs are formed dynamically by autonomous mobile peers (e.g., laptops, tablets and smart phones). These peers are connected through wireless links without an existing network infrastructure or any centralized entity and they are free to move anywhere at any time, leading to a high dynamic network topology [1]. With the rise of mobile devices and the gain in popularity of wireless networks, Internet-distributed applications like P2P file sharing have been deployed over MANETs. The latter, are commonly named as mobile P2P systems.

Indeed, these systems allow users to share and search diverse multimedia resources (e.g., audio, video, text and image data) over a MANET.

Due to the nature of MANETs, mobile P2P systems brought up many new thriving challenges, in particular with regard to the content discovery issue [2,3]. Indeed, since the network topology may change frequently, no centralized entity can be used for messages routing, thus users' queries are propagated across multiple hops in the network to locate relevant peers sharing pertinent resources. Furthermore, mobile devices in MANETs usually have limited resources such as battery power, CPU, memory capacities. Thus, the design of an efficient content discovery protocol (i.e., query routing protocol) has been of paramount importance. The thriving challenge would be to (i) locate relevant peers sharing pertinent resources for users' queries; and (ii) to ensure that those peers would be reached by considering different MANETs constraints (e.g., peer mobility, battery energy, peer load, to cite but a few).

Even though the literature witnesses a wealthy number of content discovery protocols, only few of them have considered the above-mentioned requirements [4,5,6,7,8,9]. To overcome this shortage, we introduce in this paper an efficient Content Discovery Protocol (CDP) for P2P mobile systems. The idea underlying our proposal is to route users' queries to relevant peers that are (1) more suitable to answer the query according to its content, (2) more stable to relay query hits, (3) less loaded to avoid the network congestion issue and (4) having more battery lifetime to avoid the network partitioning issue.

The remainder of this paper is organized as follows. In Section 2, we describe the pioneering existing query routing protocols (i.e., content discovery protocols) in order to show their limitations/drawbacks. In Section 3, we thoroughly describe the main idea of our content protocol, before presenting in Section 4 the simulation settings and the evaluation of the proposed CDP protocol. In addition, a comparison between CDP and another protocol is presented here. The last section concludes this study and pins down several future directions.

## 2. Literature Review

The existing content discovery protocols for P2P mobile systems can be categorized as layered or integrated protocols [10]. Layered protocols decouple functionalities of application and network layers, which enables independent development of protocols at the both layers. In this design, like in Internet, routing protocol at the application layer (i.e., content discovery protocol) operates at the top of an existing MANET routing protocol at the network layer (i.e., network routing protocol). Hence, the content discovery protocol selects the relevant logical neighbors to forward the search query to, then, it relies on an existing network routing protocol (e.g., DSR [11], AODV [12], DSDV [13], to name but a few.) to reach them. However, due to peers' mobility, these logical neighbors may not reflect the current physical topology of the ad-hoc network (i.e., underlay network). As a result, such overlay hop required by the content discovery protocol could result in a costly flooding-based route discovery by the multi-hop network routing protocol. However, MANET are a limited resource environment where the performance can be more important than portability and separation of functionalities. In this respect, integrated protocols have been proposed as alternative to the layered ones.

Within an integrated design approach, the content discovery protocol is integrated with the MANET routing protocol [10]. In [14], a decentralized and dynamic topology control protocol called TCP2P is proposed. This protocol builds a P2P overlay that closely matches the underlay network, in order to increase the fairness and provides incentive in wireless P2P file sharing applications. Although this protocol virtually controls the macroscopic usage of energy and establishes stable links between neighboring peers, it does not compromise the user satisfaction since queries are flooded regardless of their content. E-UnP2P protocol [15] builds an efficient overlay network avoiding redundant links and transmissions while ensuring connectivity among peers. It introduces a root-peer in the P2P network connecting all the other peers. Each peer maintains connection with other closer peers such that it can reach the root-peer. Using the information of its directly connected and 2-hop away (logically) neighbor peers, each peer builds up a minimum-spanning tree to identify far away peers and builds up the overlay that is closer to the physical network. Thereafter, when a peer is interested in retrieving a file, it broadcasts a lookup query in the overlay network, which takes a long time and consumes network resources extensively. In [16], the authors propose a content discovery protocol Gossiping-LB based on gossiping approach. In the Gossiping-LB, when a peer looks for forwarding a given query, it computes a forwarding probability for each of its neighbors based on its computational load (the queue utilization of the neighbor) then forwards the query to neighbors having a lower load. Indeed, this probability allows sending more messages to neighbors with lower load, while fewer messages are sent to congested peers. This protocol ensures a load balancing between peers but it floods the query regardless its content. Furthermore, it does not consider the mobility nor the battery energy factors. In [7], the authors have optimized the random walk content discovery protocol, which is a controlled flooding technique, to be more suitable for P2P mobile systems. The proposed protocol relies on a queuing system and a gradient descend technique. Indeed, relevant peers were selected according to the average queries' responses time; the average queries hit rate and the amount of energy consumption. The proposed protocol is efficient in terms of energy consumption and response time. However, it is not effective in terms of recall and precision since it does not consider the query content.

## 3. Protocol Description

The idea underlying our proposal is to introduce a Content Discovery Protocol (CDP) for mobile P2P systems as alternative to the pure query flooding protocol, which is widely used to locate resources in such systems. It is worth mentioning that within pure flooding, the query originator peer set a time-to-live counter (TTL) to a certain value then broadcasts the query to a set of $K$ random neighbors that continue to propagate it in the same way to their neighbors until the $TTL$ counter is decremented to $0$. Whenever a peer sharing pertinent resources is located, it replies by a query hits message. This message is routed back to the query originator peer through the reverse path of the query message. This method generates an excessive number of messages without ensuring that relevant peers could be located. Furthermore, query hit messages may not reach the query originator peer due to MANET environment limitations. Indeed, relay peers (i.e., peers in the reverse path of the query message) may, in the meanwhile, turn off or move out of the network at any time. To overcome these issues, in our CDP protocol peers route query messages to the $K$ most relevant neighbors instead of randomly chosen ones. To do so, we define a scoring function that allows as to select relevant neighbors that are more suitable to answer the query, more stable and less loaded to relay query hits.

In the following, we thoroughly describe the different context features used to select the $K$ relevant neighbors for a given query $q$. Thereafter, we present our neighbor selection function.

### 3.1 User's Profile and Query Content

In CDP, we consider the user's profile of neighboring peers and the content of the query in order to help the forwarding peer to locate the most relevant neighbors, which could answer the query. To achieve this goal, each peer $p_i$ maintains a user's profile for each neighbor $n_j$. The profile contains a list of the most recent past queries that the neighbor $n_j$ provided answers for. For each query it receives, the forwarding peer $p_i$ estimates the similarity $Psim$ between the query to forward q and the neighbor $n_j$ based on its profile. Roughly speaking, $Psim$ is the average similarity between the query to forward q and the past set of queries $Q_j$ that $n_j$ provided answers for. Formally,

$$Psim(n_j, q) = \frac{\sum_{q_k \in Q_j} Cosine(q_k, q)}{|Q_j|} \qquad (1)$$

Where $Cosine(q_k, q)$ is cosine similarity [14] between the queries $q_k$ and $q$. It is important to note that $Psim$ similarity stands between 0 and 1.

### 3.2 Link Stability

The link stability $l_{ij}$ between the forwarding peer $p_i$ and its neighbor $n_j$ is an important factor in the query routing process. Indeed, it is not attractive for $p_i$ to route the query $q$ to the neighbor $n_j$ if the link $l_{ij}$ between them is unstable. To compute the link stability between two peers, we define a Stablity function that combines the peer mobility and the remaining battery energy parameters. Before describing this function, we present two principle metrics. The first one takes into account the peer mobility factor to predict the lifetime of a link between two peers. The second one predicts the remaining battery time of a given peer.

Peer mobility: In MANET environment, peers are free to move from their location at any time. Hence, links between connected peers have a limited lifetime. In our protocol, we consider this important factor by predicting the lifetime of a link between the forwarding peer and its neighbors. To predict the lifetime of a link $l_{ij}$ between a peer $p_i$ and its neighbor $n_j$, we rely the affinity function $A_{ij}(t)$ defined in the RABR network protocol [18]. This function estimates at an instant $t$ the time taken by $n_j$ to move out of the range of $p_i$.

attery energy: The battery level of a mobile peer decreases when it initiates data transmission or forwards packets. A peer disconnects if the battery power finishes. This fact would undoubtedly leads to a network disconnection problem. Hence, it is not interesting to route the query message to a neighbor that has low remaining battery level $MinEnergy$ despite the lifetime between it and the forwarding peer is high. Thereby we consider this much important factor by defining the function Rtime that estimates the remaining battery time of a given neighbor. Assume that the remaining battery energy of a given neighbor $n_j$ at times $t_k$ and $t_p$ ($t_k < t_p$) are respectively $Energy(n_j, t_k)$ and $Energy(n_j, t_p)$. We define the remaining battery time $Rtime(n_j, t_p)$ of $n_j$ at time $t_p$ to reach a low energy level $MinEnergy$ as follows:

$$Rtime(n_j, t_p) = [Energy(n_j, t_p) - MinEnergy] \times \left[\frac{(t_p - t_k)}{Energy(n_j, t_k) - Energy(n_j, t_p)}\right] \qquad (2)$$

We combine the lifetime and the Rtime metrics to define the stability function $S_{ij}(t)$ that estimates at a time $t$, the time during which a peer $p_i$ could communicate directly with its neighbor $n_j$. Indeed, $S_{ij}(t)$ is defined as the minimum between the time $A_{ij}(t)$ taken by the neighbor $n_j$ to move out of the range of the forwarding peer $p_i$ and the remaining battery time $Rtime(n_j, t)$ of $n_j$ at $t$. More formally, the stability of a link $l_{ij}$ between the forwarding peer $p_i$ and its neighbor $n_j$ at $t$ is defined as follows:

$$S_{ij}(t) = Min(Rtime(n_j, t), A_{ij}(t)) \qquad (3)$$

### 3.3. Peer Load

A vital part of every network is the load-balancing factor. For instance, job completion becomes complex, if a huge load is given to the peers with less $cpu$ processing capabilities. There is a possibility of load imbalance due to the non-uniform computing/processing power of the systems. Few peers may be idle and few of them will be overloaded. A peer having a high processing power finishes its own work quickly and is estimated to stay unexploited at all most of the time. However, if we send queries only to peers that have high processing capabilities, then data packets will take routes. As a drawback, the latency is increased. With proper ways to transferring traffic load onto routes that are relatively less congested, undoubtedly would lead to a better throughput and reduced latency. In this respect, a parameter that indicates the line congestion is the queue utilization of the neighbor (i.e., Number of packets waiting in queue). Indeed, a high number of packets waiting in queue indicates line congestion. Thus, we define a Load function based on the $cpu$ capabilities and the queue utilization of the neighbor to estimate its workload. The $Load(n_j, t)$ of a neighbor $n_j$ at time $t$ is computed as follows:

$$Load(n_j, t) = \frac{1}{cpu \times (1 - u(t)) + 1} \quad (4)$$

Where $cpu$ is the processing power and $0 < u(t) < 1$ denotes the queue utilization of the neighbor $n_j$ at time $t$ (i.e., $u(t) = 1$ when the queue is full). The $Load$ metric values stand within the ]0, 1] interval. Hence, a high value of $Load$ indicates that a peer is more loaded and we would not keen to forward the query to.

### 3.4. Pertinence Function

To select the best $K$ neighbors for a given query $q$, the forwarding peer $p_i$ computes a score of each neighbor $n_j$ at time $t$ according to the following pertinence function:

$$Pertinence(n_j, q, t) = \begin{cases} MaxT \times \left( L \times Load(n_j, t) + Sim \times Psim(n_j, q) \right) & \text{if } S_{ij}(t) > MaxT \\ S_{ij}(t) \times \left( L \times Load(n_j, t) + Sim \times Psim(n_j, q) \right) & \text{otherwise} \end{cases} \quad (5)$$

Where $L$ and $Sim$ denote, respectively, the relative importance of $Load$ and $Psim$. $MaxT$ is the estimated maximum waiting time of a response from a neighbor.

Basically, in our pertinence function the link stability S factor is of paramount importance w.r.t respectively Load and $Psim$. Indeed, whenever the lifetime of the link between the sender and its neighbor is very low, then the pertinence value is also low. This case also holds even the neighbor is less loaded and is keen to answer the query (i.e., $Psim$ is high). In this case, it is not interesting to forward the query to this neighbor. Moreover, if more than two peers, having a close stability values, then the preference function enables to select peers that are less loaded and/or more similar to the query.

## 4. Performance Evaluation

In this section, we report the performance evaluation of our protocol CDP versus Gossiping-LB protocol [16]. We have chosen Gossiping-LB as baseline since it only considers the peer load factor to select relevant neighbors whereas CDP considers more than one parameters (i.e., query content, peer mobility, battery power and peer load). Worth of mention that the network simulation is carried out by the NS2 simulator [19]. Table 1 summarizes the different simulation parameters.

Table 1: Simulation Parameters

| | |
|---|---|
| TTL | The maximum number of hops that a query is allowed to travel (the horizon of the query), initialized to 3 |
| K | The maximum number of peers to be selected for a query, initialized to 3 |
| L and Sim | Represent respectively the relative importance of Load and Psim, fixed to 0.5 |
| Overlay size | Number of peers in the network, we vary the number of peers from 25 to 100 |
| MAC protocol | IEEE 802.11b |
| Area size | 500 x 500 m2 |
| Node placement | Uniform distribution |
| Node mobility | Random way-point |

### 4.1 Data Source Characteristics

The dataset used in our experiments is the Music Dataset, developed under the RARE project [20]. This dataset was obtained from a statistical analysis on Gnutella system data and from the TREC collection, which allows us to simulate our protocol in real conditions. Music Dataset is composed of 17000 documents and 700 queries.

### 4.2 Evaluation Metrics

To evaluate the routing effectiveness of our protocol, we rely on the recall and the success rate metrics, which are detailed below:

$$recall = \frac{RRD}{RLD}$$

Where, RRD denotes the number of relevant retrieved documents and RLD the number of relevant documents.

Unlike the recall metric, which assess the capacity of the system to retrieve all pertinent documents, the average success rate metric assess the capacity of the system to retrieve at least one document per query.

- Success rate: The ratio between the number of resolved queries to the total number of initiated file-lookup queries.

Besides, we are interested in assessing the routing efficiency of our protocol by computing the average file-discovery delay.

- Average file-discovery delay: The average time elapsed from the moment when a file-lookup query is sent to the moment when the replies are received.

4.3. Simulation Scenarios

Several parameters can be considered to assess the effectiveness and the efficiency of a query routing protocol for P2P mobile systems. In our case, we consider the peers velocity and the overlay size parameters. To study the effect of these parameters on our protocol, we simulate the following scenarios:

- Scenario 1: The aims of this scenario are to assess and compare the effectiveness and the efficiency of our protocol CDP against Gossiping-LB when varying the overlay size.
- Scenario 2: The aims of this scenario are to assess and compare the effectiveness and the efficiency of our protocol CDP against Gossiping-LB when varying the peers' mobility.

4.4 Results

In the remainder, we report the results of the aforementioned scenarios.

Results of Scenario 1: To study the impact of the overlay size on the retrieval effectiveness and the routing efficiency of our CDP protocol versus Gossiping-LB protocol, we vary the number of peers from 25 to 100.

Figure 1 (a) and Figure 1 (b) show that the recall and the success rate of both protocols grow as far as the number of peers rises up. Hence, the higher the number of peers is, the lower the probability to retrieve the pertinent resources is. Figure 1 (a) and Figure 1 (b) also depict that our protocol achieves better results in terms of recall and success rate than Gossiping-LB for all variations of the number of peers. Indeed, CDP increases the overall recall and success rate of Gossiping-LB by around 135 % and 70%, respectively. These encouraging results are owed to the fact that our protocol takes into account the query content, which leads to better target the adequate neighbors. However, Gossiping-LB routes the query to a less loaded neighborhood regardless the query content.

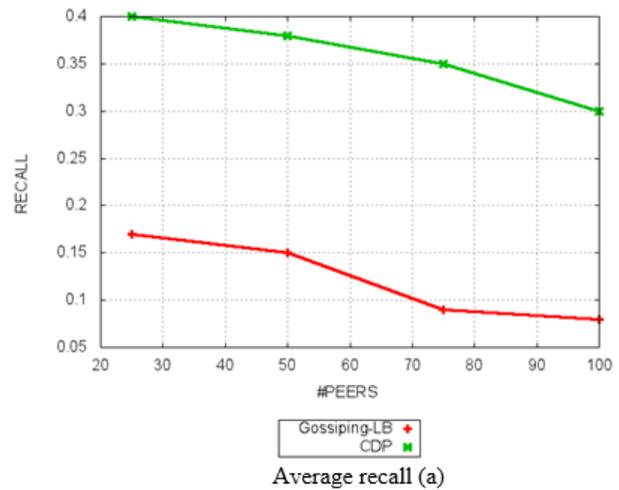

Average recall (a)

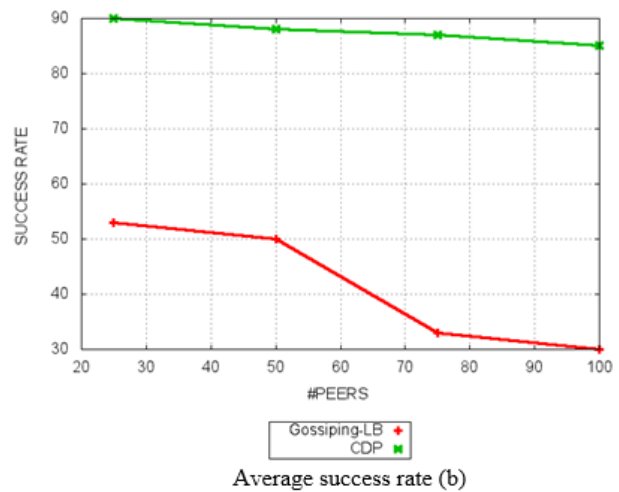

Average success rate (b)

Fig. 1 Average recall and success rate under various number of peers

Figure 2 shows that our protocol flags out shorter average-file discovery delay than Gossiping-LB at all variations of the number of peers. This is owed to the fact that our protocol relies on the historical information about past query. Such valuable information was of help to quickly locate pertinent peers reducing by the way the number of hops to reach peers sharing relevant resources. Hence, there is a few number of relays between the query originator peer and that sharing the relevant resources. Interestingly enough, Figure 2 depicts that by increasing number of peers, the average file discovery delay of both protocols rises up. This is due to the fact that the traffic overhead of both protocols increases by increasing the number of peers leading to a longer contention delay.

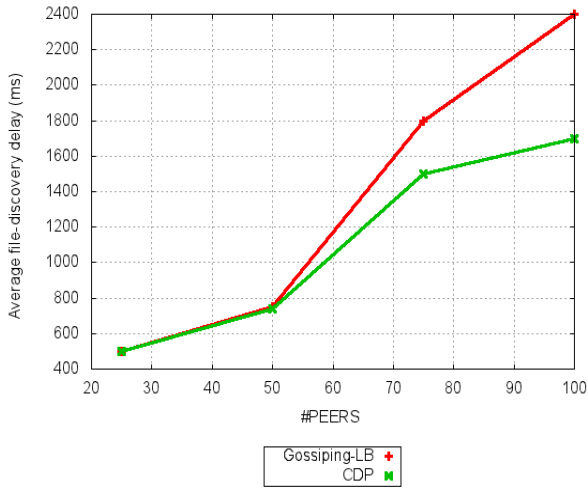

Fig. 2  Average file-discovery delay under various number of peers

Results of Scenario 2: To study the impact of the peer mobility on the retrieval effectiveness and the routing efficiency of CDP and Gossiping-LB, we vary the peers' velocity from 4.26, 7.73 to 11.68 m/s.

Figure 3 shows that the recall and the success rate of both CDP and Gossiping-LB protocols rise down by increasing the speed of peers. This can simply be explained by the fact that by increasing the speed, the links between peers become unstable. Consequently, pertinent peers are located but their answers do not reach the query initiator leading to the decrease of the average recall and success rate. Moreover, we observe that the recall and the success rate of our protocol are higher than those of Gossiping-LB for all the variations of the speed. We also remark that by considering node mobility constraint our protocol is more resistant to speed increase than Gossiping-LB.

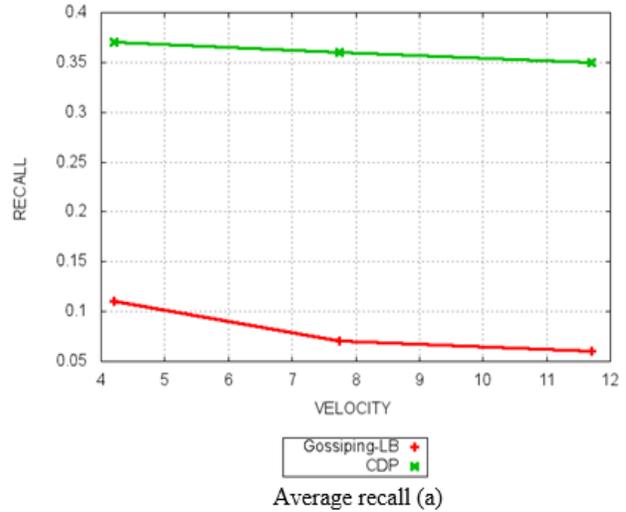

Average recall (a)

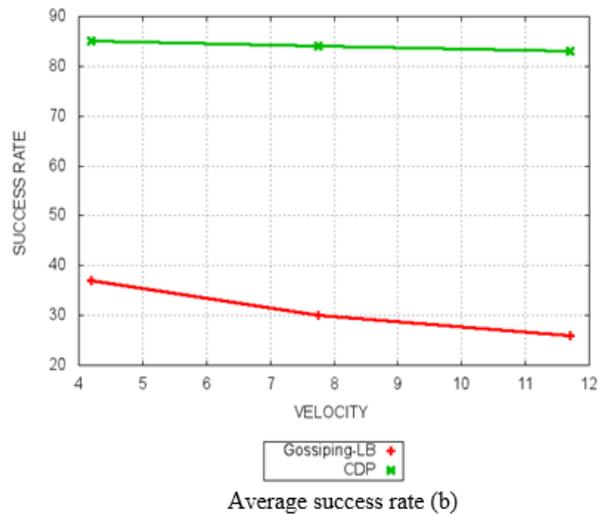

Average success rate (b)

Fig. 3  Average recall and success rate under various peers' speed

Figure 4 shows that our protocol has shorter average-file discovery delay than Gossiping-LB according to various peers' velocity. Indeed, CDP achieves 25% less file-discovery delay than Gossiping-LB. This is obtained thanks to the fact that CDP considers the node mobility constraint whereas Gossiping-LB considers only peer's load factor.

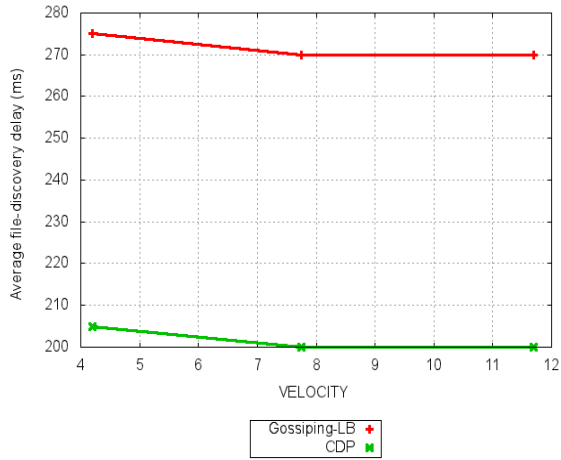

Fig. 4  Average file-discovery delay under various peers' speed

## 5. Conclusion and Future Works

In this paper, we have introduced a content discovery protocol for unstructured mobile P2P systems. Our protocol learns the user's behavior, in order to route the forthcoming queries to neighbors keen to provide adequate answers with regard to the query content. Furthermore, it considers mobility and battery energy of relay peers to ensure that relevant peers would be reached. It also takes into account the peer load factor to avoid latency.  The experimental results have highlighted that our proposed protocol has performed better than pioneering protocol in the literature. Obvious pointers for future work include highlighting certain shortcomings in the evaluation of our protocol. Thus, further evaluation is ongoing.

**Acknowledgments**

The authors gratefully acknowledge the approval and the support of this research study by the grant no. 7286-SCI-2017-1-8-F from the Deanship of Scientific Research at Northern Border University, Arar, K.S.A.

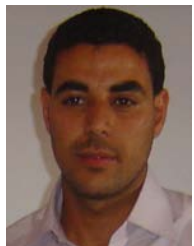
**Taoufik Yeferny** received his M.C.S. and Ph.D. degrees in Computer Sciences from the University of Tunis El Manar, Tunisia, in 2009 and 2014, respectively. From 2013 to 2016, he was an Assistant Professor at the High Institute of Applied Languages and Computer Science of Beja, Tunisia. Since 2016, he has been an Assistant Professor at the Northern Border University, KSA. His current research interests include mobile P2P systems, Vehicle Ad hoc Network and Intelligent Transportation Systems.

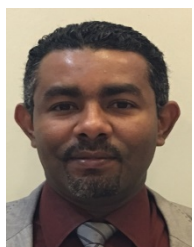
**Sofian Hamad** received his B.Sc. (Honors) degree in computer science from the Future University, Khartoum, Sudan, in 2003. In 2007, he received an M.Sc. in management business administration (MBA) from the Sudan Academy of Science. In 2013, he received a Ph.D. in electrical engineering from Brunel University in London, the United Kingdom. Dr. Hamad is currently an assistant professor in the Department of Computer Science at Northern Border University. His current research interests include Ad-hoc and Mesh wireless networks, Vehicular Technology, and the Internet of Things.

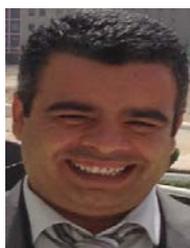
**Salem Belhaj** was born in Tunis on June 10, 1977. He received his degree in computer engineering from the National School of Computer Sciences of Tunis in 2002, an M.S. from the University of Versailles in 2003, and his Ph.D. in Computer Science from the National School of Computer Sciences of Tunis. Dr. Belhaj is currently an Assistant Professor in the Department of Computer Science at Northern Border University. His research interests include routing in sensor networks and IoT, network delay modeling, analysis, prediction, delay-based congestion control, and teleoperation.